# Comment on "Superconducting order parameter in partially substituted Bi2212 single crystals as measured by the tunneling effect"

Hancotte *et al*. [1] have presented tunneling measurements performed on pure, Ni- and Zn-doped $Bi_2Sr_2CaCu_2O_{8+x}$ (Bi2212) single crystals. One may wonder why tunneling data [2] obtained on same set of Ni- and Zn-doped Bi2212 single crystals are different (see, for example, Fig. 3(b) of Ref. 1 and Fig. 3(b) of Ref. 2) [3]. Moreover, these two sets of measurements have been performed on same measurement set up.

The main conclusions in Ref. 1 are (excluding the speculations):
(i) the distribution of the magnitude of tunneling gap in pure Bi2212 single crystals ($T_c \sim 87$ K) varies between 42.5 and 65 meV;
(ii) the distribution of the magnitude of tunneling gap in Ni-doped Bi2212 single crystals ($T_c \sim 75$ K) varies between 50 and 108 meV;
(iii) in a Zn-doped Bi2212 single crystal ($T_c = 79$ K), "a light decrease of the energy gap is observed in this particular case but its values still within the range of gap values usually measured on pure crystals [see Fig. 3(a)] and no increase of this value was observed on the small number of measured Zn-substituted samples."
(iv) the temperature dependencies of pure and N-doped Bi2212 single crystals lie somehow above the BCS temperature dependence.

In Ref. 2, the distribution of the magnitude of tunneling gap in Ni- and Zn-doped Bi2212 is different and varies between 25 and 115 meV and between 40 and 115 meV, respectively. Why is this? First of all, one has to note that the data in Fig. 3(b) of Ref. 1 and in Fig. 3(b) of Ref. 2 have different meaning: in Ref. 1, Figure 3(b) shows the distribution of tunneling gap in different samples while Figure 3(b) in Ref. 2 presents the distribution of tunneling gap in *one* sample. However, it doesn't explain the reason for the discrepancy.

The main reason for the discrepancy between the data presented in Refs. 1 and 2 is not scientific but rather prosaic. Assuming the magnetic origin of the superconducting mechanism and the BCS temperature dependence in copper-oxides a Professor during a discussion with his Ph.D. student [4] said that the tunneling gap has to increase in Ni-doped Bi2212 and to decrease in Zn-doped samples. Later, the Ph.D. student showed to the Professor only a part of measured data obtained on Ni- and Zn-doped Bi2212 single crystals, which correspond to the statement of the Professor. The data have been published [1]. Then, the Ph.D. student happily got a job in the largest insurance company one month before his Ph.D. defense and left the university. The Professor never had a possibility to look in student's files with raw data. Fortunately, the author of this note had such possibility.

The raw data of the distribution of the magnitude of tunneling gap in Ni-doped Bi2212 were artificially "cut" from both sides. The result of this action is presented in Fig. 3(b) of Ref. 1. About 80% of measured temperature dependencies on pure and Ni-doped Bi2212 single crystals lie below the BCS temperature dependence (see, for example, Ref. 5 and 6). Measurements on a few Zn-doped samples have been performed in a harry without going into details. However, all data presented in Ref. 1 concerning the pure Bi2212 single crystals (with the exception for the temperature dependence of tunneling gap) correspond to measured results.

Thus, basically, there is no difference in the distributions of the magnitude of tunneling gap for the Ni- and Zn-doped Bi2212 single crystals as it is shown in Figs. 3(a) and 3(b) of Ref. 2.

A. Mourachkine

Université Libre de Bruxelles,
Service de Physique des Solides, CP233,
Boulevard du Triomphe,
B-1050 Brussels,
Belgium

[1] H. Hancotte *et al.*, Phys. Rev. B **55**, R3410 (1997).
[2] A. Mourachkine, to be published in J. of Superconductivity (cond-mat/9902355).
[3] Initially, this Comment was written for a referee in one of the journals.
[4] The names will be omitted.
[5] S. I. Vedeneev *et al*., JETP Lett. **47**, 679 (1988).
[6] A. Mourachkine, (unpublished) cond-mat/9901282.